\documentclass[twocolumn,aps,prl,showpacs]{revtex4}

\usepackage{graphicx}
\begin{document}

\title{Time scale for the onset of Fickian diffusion in supercooled liquids}
\author{Grzegorz Szamel and Elijah Flenner}
\affiliation{Department of Chemistry,
Colorado State University, Fort Collins, CO 80525}

\date{\today}


\begin{abstract}
We propose a quantitative measure of a time scale on which Fickian diffusion
sets in for supercooled liquids and use Brownian Dynamics computer 
simulations to determine the temperature dependence of this onset time in 
a Lennard-Jones binary mixture. The time for the onset of 
Fickian diffusion ranges between 6.5 and 31 times the $\alpha$ relaxation 
time (the $\alpha$ relaxation time is 
the characteristic relaxation time of the incoherent 
intermediate scattering function). The onset time 
increases faster with decreasing temperature than the $\alpha$ 
relaxation time. Mean squared displacement at the onset time 
increases with decreasing temperature.
\end{abstract}
\maketitle

Understanding the origin of the extreme slowing down of liquids' 
dynamics upon approaching the
glass transition and the nature of the transition itself has 
been of great interest for several decades. 
A lot of recent activity has been stimulated by 
the recognition that close to the transition the liquids' dynamics 
become not only very sluggish but also increasingly 
heterogeneous \cite{reviews}. 
While the presence of dynamic heterogeneities is commonly accepted, 
the details of their spatial and temporal structure have been only
partially established. In particular, the question of the lifetime
of dynamic heterogeneities is quite controversial: whereas two 
separate experiments \cite{Ediger12,vandenBout} 
found that near the glass transition the lifetime is significantly longer
than the $\alpha$ relaxation time, other experimental studies 
\cite{NMR,solvation,diel} found the 
lifetime to be comparable to the $\alpha$ relaxation time. In principle, 
the controversy can be resolved by postulating 
that the temperature dependence of the lifetime is stronger than that 
of the $\alpha$ relaxation time \cite{EdigerAR}. However, the physical
interpretation of the new time scale remains unclear. 
On the computational side, there have been 
a few attempts to estimate the lifetime of dynamic heterogeneities.
Most of them \cite{Perera,DoliwaJNCS,FS,YGC} found the lifetime
to be comparable to the $\alpha$ relaxation time. 
To the best of our knowledge, the only exception is a recent preprint 
\cite{Berthierlife} in which it is shown that, in a kinetically constrained 
spin model resembling a fragile glass former, the lifetime is a few times 
longer than the $\alpha$ relaxation time.  More importantly, 
Ref. \cite{Berthierlife} found that the lifetime increases 
with decreasing temperature somewhat faster than 
the $\alpha$ relaxation time. It should be noted that one earlier 
study \cite{YO} also found that at a low temperature the lifetime
of dynamic heterogeneities is a few times longer than the $\alpha$ 
relaxation time. However, a careful study of the temperature 
dependence of these two times have not been performed. 
Thus, the question of the existence of 
a time scale longer \emph{and} increasing faster than the $\alpha$ relaxation 
time remains unresolved.

The goal of our study is to investigate the temperature dependence of 
a different characteristic time that is related to the lifetime 
of dynamic heterogeneities: the time for the onset of Fickian diffusion.

In order to define the onset time we use, as an indicator of Fickian
diffusion, 
the probability distribution of the logarithm of single-particle
displacement, $\log_{10}(\delta r)$,
during time $t$, $P(\log_{10}(\delta r);t)$ 
\cite{Puertas,Reichman,Cates,FS2}. This distribution is defined in such a way
that the integral $\int_{x_0}^{x_1} P(x;t) \mbox{d}x$ is the
fraction of particles whose value of $\log_{10}(\delta r)$ is between
$x_0$ and $x_1$. The probability distribution
$P(\log_{10}(\delta r);t)$ can be obtained from the self part of the 
van Hove correlation function  \cite{HansenMcDonald},
$P(\log_{10}(\delta r);t) = \ln(10) 4 \pi \delta r^3 G_s(\delta r,t)$.
The probability distribution $P(\log_{10}(\delta r);t)$ 
is a convenient indicator of Fickian diffusion 
because if particles move via Fickian diffusion
then the self part of the van Hove function is Gaussian and the 
shape of the probability distribution $P(\log_{10}(\delta r);t)$ 
is independent of time. In particular, the height of the peak of 
this distribution is equal to 
$\ln(10) \sqrt{54/ \pi}\;e^{-3/2} \approx 2.13$ and 
deviations from this value indicate non-Fickian particle motion.
We define the time for the onset of Fickian diffusion, $\tau_F$, as the time
at which the peak of $P(\log_{10}(\delta r);t)$ is equal to 
90\% of its value for a Gaussian distribution of displacements,
$P(\log_{10}(\delta r_{max});\tau_F) \approx 1.92$. 
We will discuss the threshold value of $90\%$ 
together with a different indicator of Fickian diffusion at the end of
this Letter.

It should be noted that a deviation of the probability distribution
$P(\log_{10}(\delta r);t)$
from its universal shape expected for Fickian diffusion indicates
dynamic heterogeneity. However, in principle, the inverse is not necessarily
true. Thus, the time for the onset of Fickian diffusion is probably
only a lower bound for the lifetime of dynamic heterogeneities.

To investigate the onset time we use the trajectories generated
by an extensive Brownian Dynamics simulation study of a 80:20 
Lennard-Jones binary mixture introduced by Kob and Andersen \cite{KobAndersen}.
Briefly, the potential is given by
$V_{\alpha \beta} =  4\epsilon_{\alpha \beta}
\left[ (\sigma_{\alpha \beta}/r)^{12} - (\sigma_{\alpha \beta}/r)^6 \right]$,
where $\alpha,\beta \in \{A,B\}$, and $\epsilon_{AA} = 1.0$, $\epsilon_{AB} 
= 1.5$,
$\epsilon_{BB} = 0.5$, $\sigma_{AA} = 1.0$, $\sigma_{AB} = 0.8$, 
and $\sigma_{BB} = 0.88$
(all the results are presented in reduced units where $\sigma_{AA}$ 
and $\epsilon_{AA}$ are the units of length and energy, respectively).
A total of $N = 1000$ particles were simulated with a 
fixed cubic box length of 9.4.
The details of this study have been presented 
elsewhere \cite{SzamelFlenner,FS2}.
In the present investigation 
we use only some of the temperatures simulated before: 
$T=1.0$, 0.9, 0.8, 0.6, 0.55, 0.5, 0.47, and 0.45.
The previous runs at the temperature $T=0.45$ have 
been extended by 60\%; the A particles' mean squared displacement 
at the end of the extended runs is about 16.
We present the results for the A particles only. The 
results for the B particles are qualitatively the same, although 
the statistics is worse due to the smaller number of B particles.
The temperature dependence is presented by plotting various 
quantities \textit{vs.} 
$T-T_c$ where $T_c=0.435$ is the
crossover temperature \cite{KobAndersen,FS2}. 
This is a convenient way to expand the temperature scale
and it should not imply an endorsement of any particular theoretical 
approach.

\begin{figure}
        \includegraphics[scale=0.25]{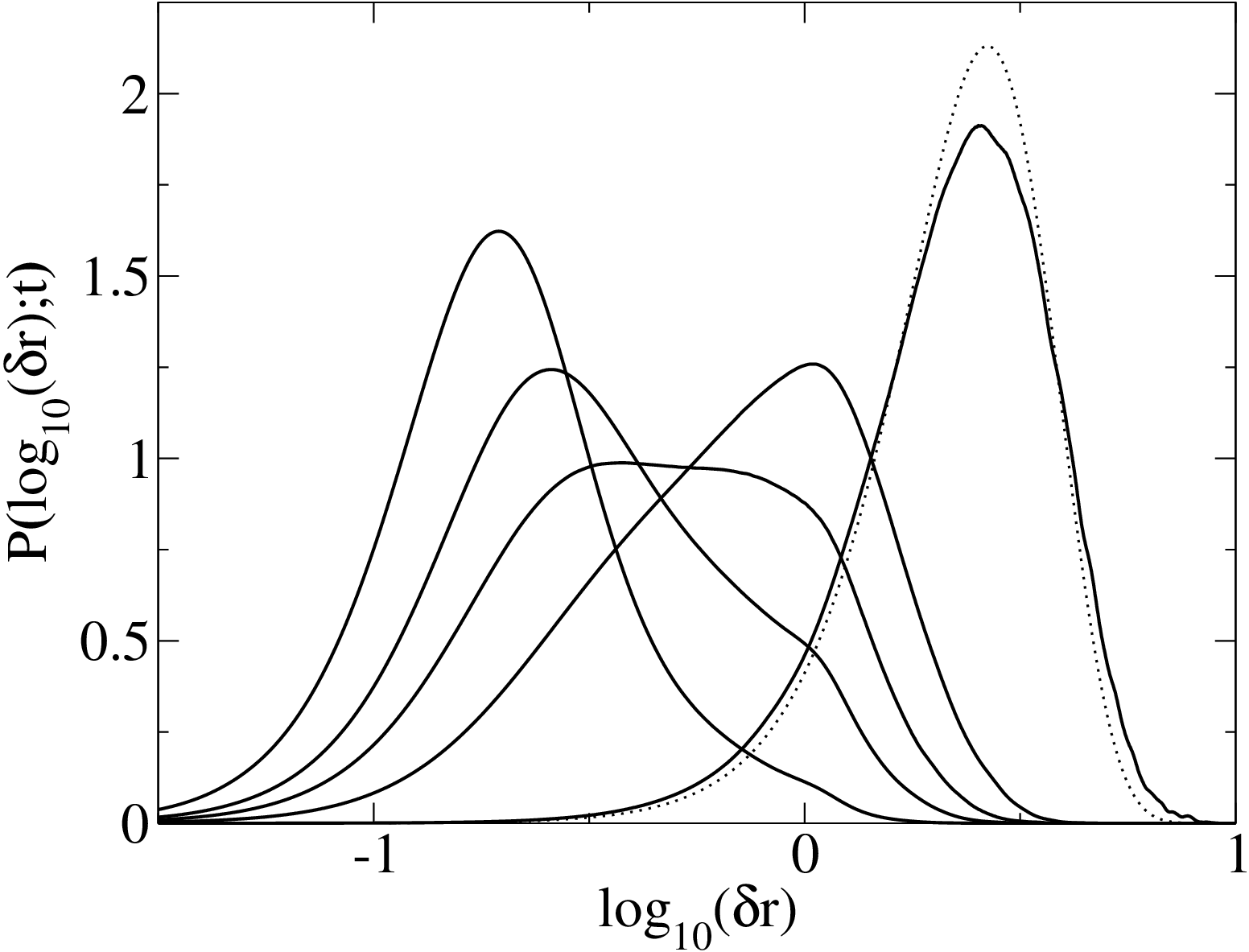}
\caption{\label{Plog45}The probability of the logarithm of single particle
displacements $P(\log_{10}(\delta r);t)$ at $T=0.45$ for the A particles. 
The time $t$ is equal to, from left to right, $\tau_{ng}=160$,
$\tau_{\alpha}=725$, $\tau_{nng}=1405$, $\tau_{0.1}=2990$, and 
$\tau_F=22391$ (see text for definition of these times).
For a comparison we also show, as a dotted line, 
$P(\log_{10}(\delta r);t)$ resulting from a Gaussian distribution 
of displacements.}
\end{figure}

We start by showing in Fig. \ref{Plog45} the probability distributions 
$P(\log_{10}(\delta r);t)$ at $T=0.45$ for the A particles 
at several times characteristic for the relaxation
of the system. The first one is the time at which
the non-Gaussian parameter 
$\alpha_2(t) = \frac{3}{5}\left<\delta r^4 \right>
/\left<\delta r^2\right>^2 - 1$
reaches the maximum value, $\tau_{ng}$.
The second one is the $\alpha$ relaxation time, $\tau_{\alpha}$, which is
defined in the usual way: $\tau_{\alpha}$ is the time at which 
the incoherent intermediate scattering function for a wave vector near
the peak of the static structure factor is equal to $1/e$ of its initial 
time value, $F_s(k;\tau_{\alpha})=1/e$. The third time is  the time 
at which a new non-Gaussian parameter 
$\gamma(t) = \frac{1}{3} \left< \delta r^2 \right> 
\left< 1/\delta r^2 \right> - 1$ \cite{FS2} 
reaches the maximum value, $\tau_{nng}$. We argued in Ref. 
\cite{FS2} that deviations 
of $P(\log_{10}(\delta r);t)$ from its Fickian shape are most evident
for times comparable to $\tau_{nng}$. 
The fourth time is the time at which 
the incoherent intermediate scattering function for a wave vector near
the peak of the static structure factor is equal to 10\% of its initial 
time value, $F_s(k;\tau_{0.1})=0.1$.
The final time is the onset time, 
$\tau_F$, \textit{i.e.} the time at which
the peak of $P(\log_{10}(\delta r);t)$ is equal to 
the 90\% of its value for a Gaussian distribution of displacements. 
For a comparison we also show 
a $P(\log_{10}(\delta r);t)$ resulting from a Gaussian distribution 
of displacements. It is clear from Fig. \ref{Plog45} that at 
shorter times, \textit{i.e.} at $\tau_{ng}$, $\tau_{\alpha}$
$\tau_{nng}$, and $\tau_{0.1}$ 
the probability distributions $P(\log_{10}(\delta r);t)$ 
deviate strongly from the shape resulting from a Gaussian
distribution of displacements. While there are still noticeable 
differences even at $\tau_F$, we believe that these are small enough
to consider $\tau_F$ the onset time for Fickian diffusion.

\begin{figure}
        \includegraphics[scale=0.25]{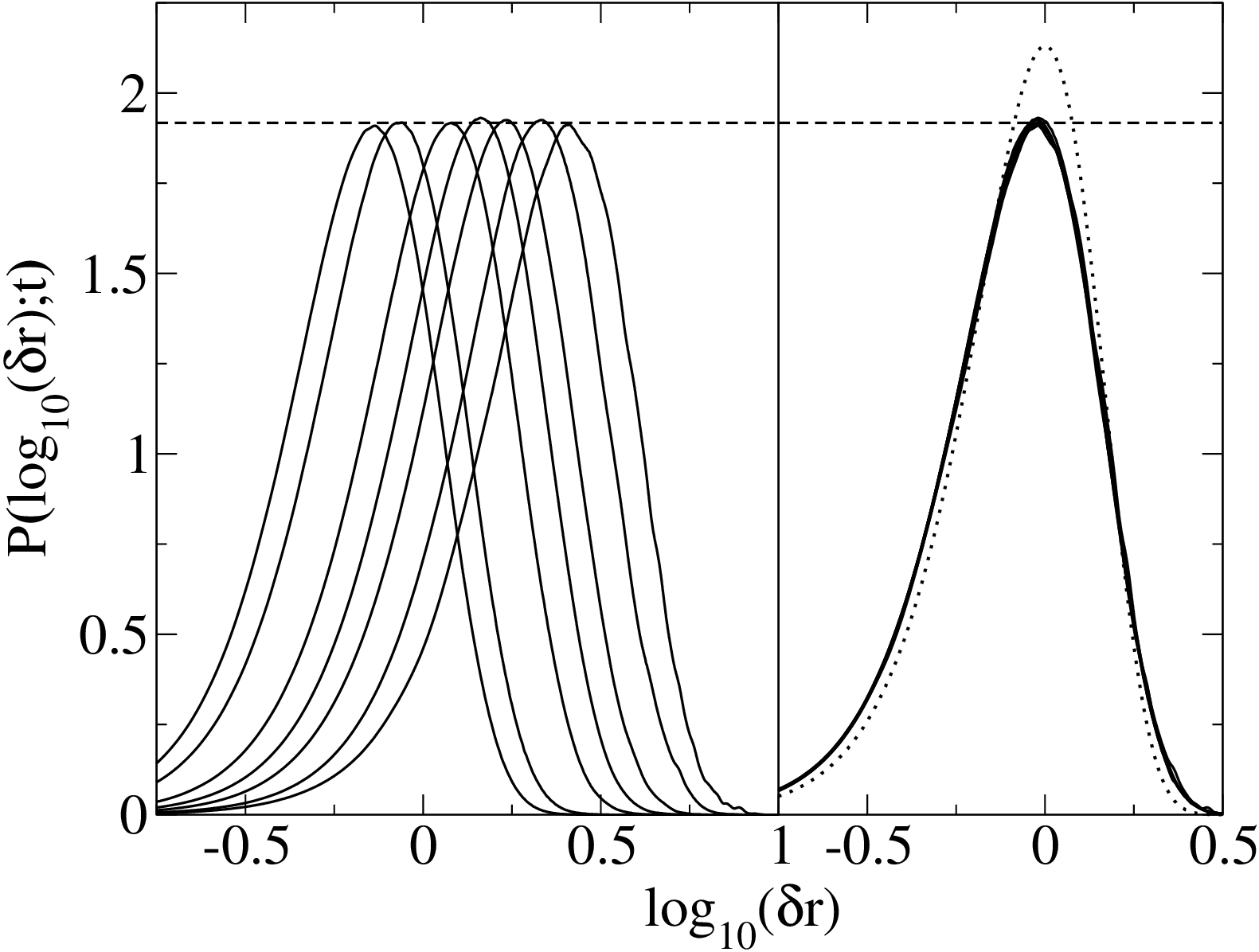}
\caption{\label{Plog} Left panel: 
The probability of the logarithm of single particle
displacements $P(\log_{10}(\delta r);\tau_F)$ for the A particles 
for $T = 1.0$, 0.8, 0.6, 0.55, 0.50, 0.47, 0.45, listed from
left to right. Right panel: the probability distributions from the
left panel shifted in such a way that $\left<\delta r^2\right>=1$;
for a comparison we also show, as a dotted line,
$P(\log_{10}(\delta r);t)$ resulting from a Gaussian distribution
of displacements with $\left<\delta r^2\right>=1$.}
\end{figure}

In Fig. \ref{Plog} we show $P(\log_{10}(\delta r);\tau_F)$ 
for the A particles for $T = 1.0$, 0.8, 0.6, 0.55, 0.50, 0.47, and 0.45.
It should be noted that with decreasing temperature the probability
distributions at $\tau_F$ shift toward larger displacements.
In other words, mean squared displacement at the onset of Fickian
diffusion increases with decreasing temperature. The right panel
indicates that the shape of $P(\log_{10}(\delta r);\tau_F)$
is temperature-independent and, therefore, the late-time liquids'
dynamics are, up to rescaling of the time and distance scales, similar.

\begin{figure}
        \includegraphics[scale=0.25]{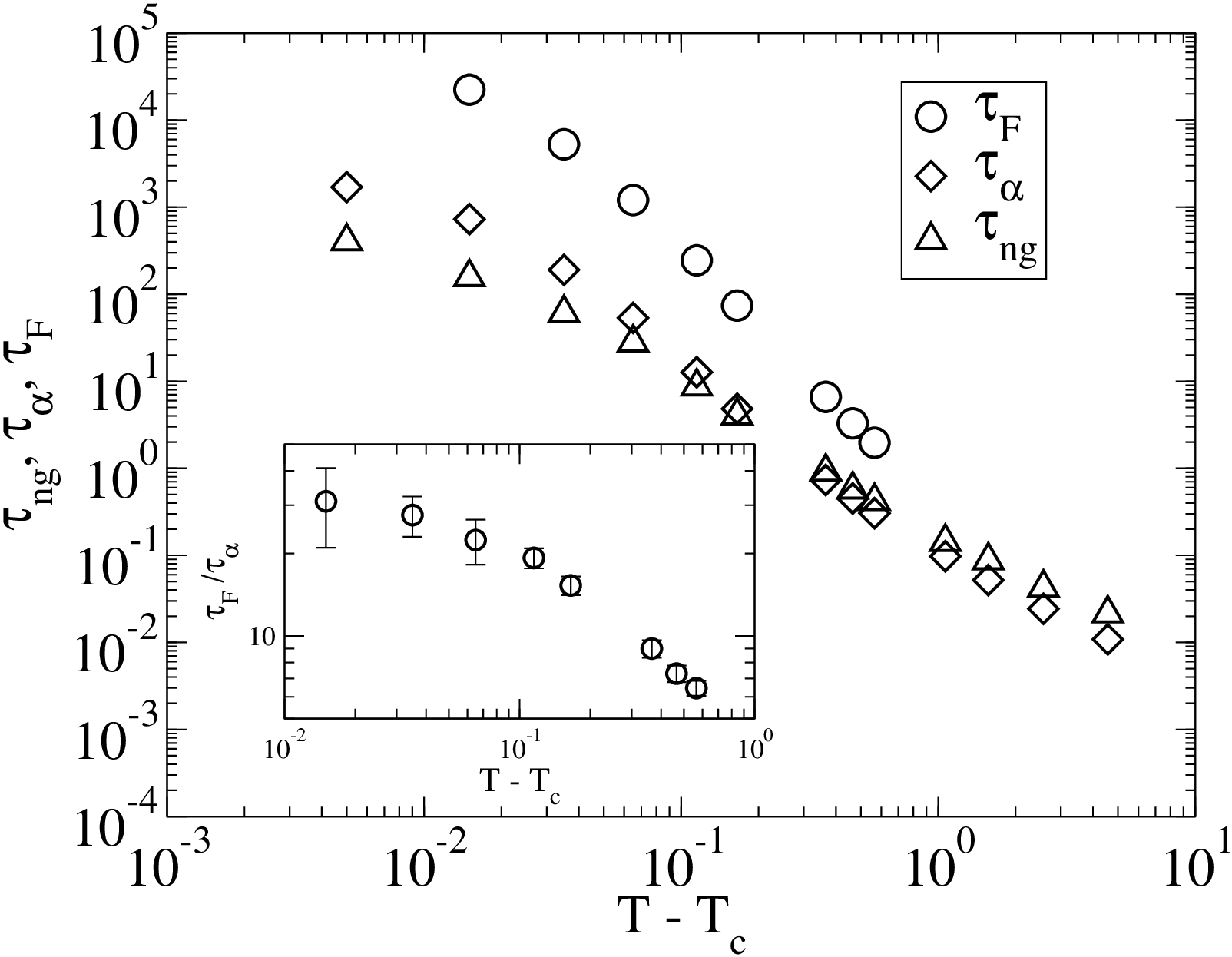}
\caption{\label{times} Temperature dependence of characteristic
times for the A particles. Triangles: the time at which
the non-Gaussian parameter $\alpha_2(t)$ 
reaches the maximum value, $\tau_{ng}$.
Diamonds: the $\alpha$ relaxation time, $\tau_{\alpha}$.
Circles: the onset time for Fickian diffusion, $\tau_F$. Inset: 
temperature dependence of the ratio $\tau_F/\tau_{\alpha}$.}
\end{figure}

Fig. \ref{times} presents our main result: comparison of 
the temperature dependence of 
of the onset time for Fickian diffusion, $\tau_F$, with that
of other characteristic times. We find that, in the temperature range
considered in this Letter, the onset time is between 6.5 $\tau_{\alpha}$ 
and 31 $\tau_{\alpha}$. More importantly, the ratio
of the onset time and the $\alpha$ relaxation time grows 
with decreasing temperature. Interestingly, the temperature dependence
of this ratio becomes somewhat weaker with decreasing temperature:
it appears stronger in the range $0.1 \le T-T_c \le 1$
than in the lower temperature range $0.01 \le T-T_c \le 0.1$. 

\begin{figure}
        \includegraphics[scale=0.25]{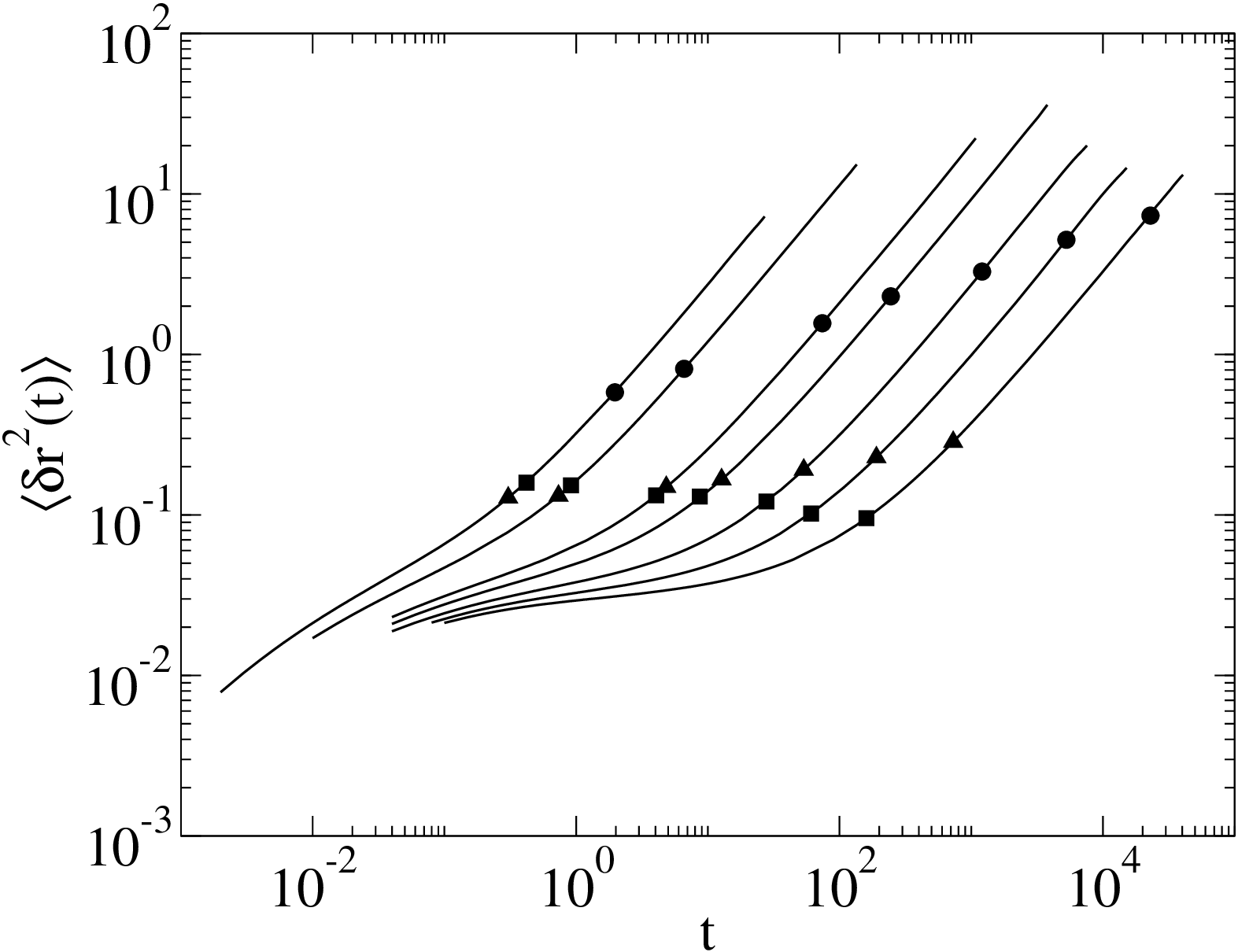}
\caption{\label{msdb} The time dependence of the 
mean square displacement for the A particles
for $T = 1.0$, 0.8, 0.6, 0.55, 0.50, 0.47, and 0.45 listed from
left to right. The symbols are placed at different characteristic
times. Squares: the time at which
the standard non-Gaussian parameter reaches the maximum value, $\tau_{ng}$.
Triangles: the $\alpha$ relaxation time, $\tau_{\alpha}$.
Circles: the onset time for Fickian diffusion, $\tau_F$.}
\end{figure}

In Fig. \ref{msdb} we place the results shown in Figs. \ref{Plog}
and \ref{times} in the context of the time dependence of the
mean squared displacement. On the time scale of 
$\tau_{ng}$ the mean squared displacement
has not yet reached the linear dependence on time and thus the diffusion
is obviously non-Fickian. Moreover, on the time scale of $\tau_{\alpha}$ 
the mean squared displacement is, at most, at the borderline
of the linear time dependence. On the other hand, the onset time,
$\tau_F$, occurs well within the regime of apparent 
linear time dependence of the mean squared displacement.
Note that there is an important practical message from 
Fig. \ref{msdb}: if one monitors only the time-dependent mean square 
displacement, one can significantly underestimate
the length of the run necessary to achieve Fickian diffusion. 

\begin{figure}
        \includegraphics[scale=0.25]{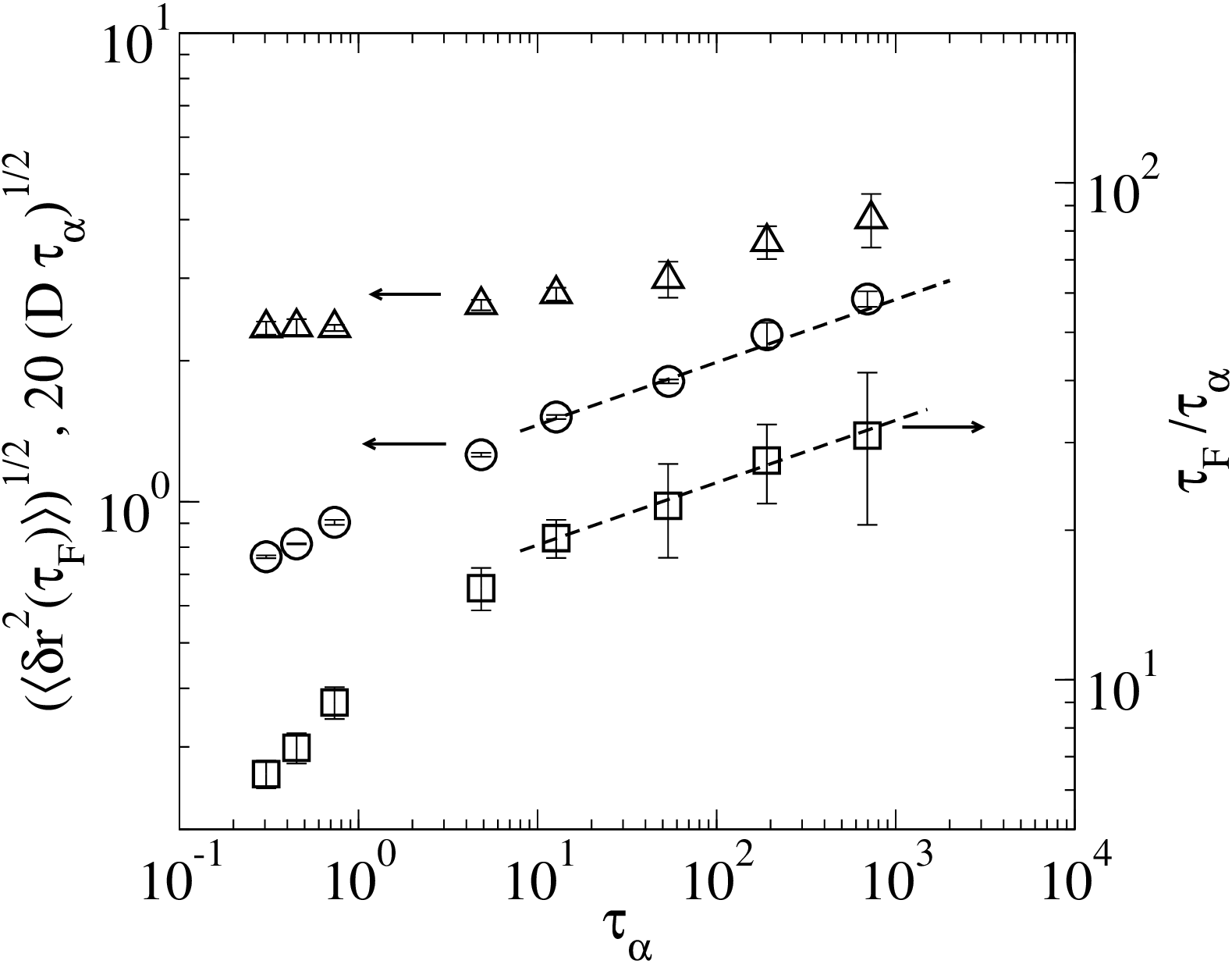}
\caption{\label{msdntau} Circles (left vertical axis): 
the root mean squared displacement at the onset time, 
$(\left<\delta r^2 (\tau_F) \right>)^{1/2}$. Triangles (left vertical axis): 
the square root of the product of the self-diffusion
coefficient and the $\alpha$ relaxation time (multiplied by 20 for 
convenience), $20 \left(D \tau_{\alpha}\right)^{1/2}$. Squares (right
vertical axis): the ratio of 
the onset time and the $\alpha$ relaxation time, $\tau_F/\tau_{\alpha}$. 
All quantities are plotted \textit{vs.}
the $\alpha$ relaxation time; all the data pertain to the A particles.
Dashed lines indicate scaling relationships
$\left(\left<\delta r^2(\tau_F) \right>\right)^{1/2} 
\propto \tau_{\alpha}^{0.13}$
and $\tau_F/\tau_{\alpha} \propto \tau_{\alpha}^{0.13}$.}
\end{figure}

Having identified the Fickian crossover time, $\tau_F$, we can
define a characteristic length scale, the root mean squared displacement
at the onset time, $\left(\left<\delta r^2 (\tau_F) \right>\right)^{1/2}$.
It follows from Figs. \ref{Plog} and \ref{msdb} that this length increases 
with decreasing temperature. Our characteristic length should be related
to the so-called Fickian crossover length $l^*$ introduced by
Berthier \textit{et al.} \cite{BCG} The latter 
length was defined through 
the wavevector dependence of the relaxation time of a supercooled liquid. 
Roughly speaking, $l^*$ is the length scale on which diffusion is
Fickian on all time scales. The prediction of Ref. \cite{BCG} 
was that this length scale changes with temperature as 
the square root of the product 
of the self-diffusion coefficient and the $\alpha$ relaxation time, 
$l^* \propto \left(D \tau_{\alpha}\right)^{1/2}$ \cite{KSKumar}. 
In Fig. \ref{msdntau} we compare temperature dependence 
of $\left(\left<\delta r^2 (\tau_F) \right>\right)^{1/2}$ to that of
of $l^*$ (note that we plot these length scales \textit{vs.} the $\alpha$ 
relaxation time; this is in the spirit of Refs. 
\cite{BCG,otherC} in which it is argued that the glass transition 
is a manifestation of a zero-temperature critical point).
The root mean squared displacement at the onset time grows 
with increasing $\tau_{\alpha}$ and at the lowest temperatures 
there is an apparent scaling relationship,
$\left(\left<\delta r^2 (\tau_F) \right>\right)^{1/2} 
\propto \tau_{\alpha}^{0.13}$. In contrast, 
$\left(D \tau_{\alpha}\right)^{1/2}$ is initially temperature-independent. 
This is due to the fact that the Stokes-Einstein relation is violated
only for $\tau_{\alpha} > 1$ (\textit{i.e.} for $T<0.8$) \cite{FS3}. However, 
at longer $\alpha$ relaxation times (\textit{i.e.} at lower
temperatures) $\left(D \tau_{\alpha}\right)^{1/2}$
has a temperature dependence similar to that of 
$\left(\left<\delta r^2 (\tau_F) \right>\right)^{1/2}$. 
Finally, we show in Fig. \ref{msdntau} that at longer $\alpha$ 
relaxation times (\textit{i.e.} at lower temperatures)
the ratio 
$\tau_F/\tau_{\alpha}$ appears to grow with increasing 
$\alpha$ relaxation time as 
$\tau_F/\tau_{\alpha} \propto \tau_{\alpha}^{0.13}$. 
It is not clear whether the scaling relations indicated in Fig. \ref{msdntau}
have any deeper significance.
It could be even argued that if they
continue for another 7 or 8 orders of magnitude of $\tau_{\alpha}$ 
(\textit{i.e.} up to $\tau_{\alpha}$ 
comparable to that at the laboratory glass transition temperature), 
the resulting $\tau_F$ would be greater than the longest 
experimentally observed heterogeneity lifetime.

To summarize, we proposed a quantitative definition of the onset
time for Fickian diffusion and investigated its temperature dependence
in a Lennard-Jones binary mixture. We found that the onset time
is considerably longer than the $\alpha$ relaxation time and,
more importantly, it increases faster with decreasing temperature than
the $\alpha$ relaxation time. Our definition of the onset time
relies upon one particular indicator of Fickian diffusion, 
the probability distribution of the logarithm of single-particle
displacement, $P(\log_{10}(\delta r);t)$, and upon adopting a particular
numerical criterion for the onset of Fickian diffusion,
peak height equal to the 90\% of its Fickian value. 
This procedure seems reasonable in that it results in
non-Fickian motion being present only at temperatures at and below
$T\approx 1.0$. This temperature has been identified before as so-called 
onset temperature for slow dynamics \cite{onset}. To test the robustness
of our main result we also tried using a different indicator of
Fickian diffusion: the new non-Gaussian parameter that we introduced
recently \cite{FS2}. In this approach we defined the onset time
for Fickian diffusion to be the time at which the new non-Gaussian
parameter is equal to $1/3$. This particular numerical value results 
in non-Fickian motion being present only at temperatures at and below
$T\approx 0.8$ \cite{commentSE}. 
The resulting onset times are somewhat shorter than
the ones presented in this Letter. However, the temperature dependence
of the onset time defined using the new non-Gaussian parameter 
is similar to that of the onset time defined using $P(\log_{10}(\delta r);t)$.
More interestingly, we found that the shapes of  $P(\log_{10}(\delta r);t)$
at the onset times defined using the new non-Gaussian parameter 
are very similar and, in particular, the height of the peak is approximately
temperature independent and equal to 85\% of its value for Fickian 
diffusion. 

Finally, we would like to point out that the results presented here violate
the time-temperature superposition principle: in order to superimpose 
the probability distributions $P(\log_{10}(\delta r);t)$ shown
in the left panel of Fig. \ref{Plog} we have to shift $\log_{10}(\delta r)$
by $\log_{10}\left(\left<\delta r^2 (\tau_F) \right>\right)^{1/2}$. 
The more usual shift procedure, agreeing with the time-temperature
superposition, would involve the $\alpha$ relaxation time rather
than the onset time $\tau_F$ which has temperature dependence different from 
$\tau_{\alpha}$. 

G.S. thanks Mark Ediger for many discussions on dynamic
heterogeneity experiments that stimulated this work. 
We gratefully acknowledge the support of NSF Grant No.~CHE 0111152.


\begin{thebibliography}{99}
\bibitem{reviews} Experimental studies of dynamic heterogeneities 
have been reviewed in Refs. \cite{EdigerAR,RichertJPCM}. 
For a very recent review of simulational investigations see Ref. 
\cite{AndersenPNAS}.
\bibitem{commentSE}  Note that according to both definitions the appearance of
the non-Fickian motion correlates with violation of the 
Stokes-Einstein relation.
\end{thebibliography}
\end{document}